\documentstyle[11pt]{article}

\let\a=\alpha \let\b=\beta

\let\la=\label  
  
\def\nn{\nonumber} \def\bd{\begin{document}} \def\ed{\end{document}}
\def\ds{\documentstyle} \let\fr=\frac \let\bl=\bigl \let\br=\bigr
\let\Br=\Bigr \let\Bl=\Bigl
\let\bm=\bibitem
\let\na=\nabla
\let\pa=\partial \let\ov=\overline
\newcommand{\be}{\begin{equation}}
\newcommand{\ee}{\end{equation}}
\def\ba{\begin{array}}
\def\ea{\end{array}}
\def\ft#1#2{{\textstyle{{\scriptstyle #1}\over {\scriptstyle #2}}}}
\def\fft#1#2{{#1 \over #2}}
\def\del{\partial}
\def\vp{\varphi}
\def\sst#1{{\scriptscriptstyle #1}}
\def\wdg{{\,\sst \wedge\,}}
\def\oneone{\rlap 1\mkern4mu{\rm l}}
\def\td{\tilde}
\def\wtd{\widetilde}
\newcommand{\ho}[1]{$\, ^{#1}$}
\newcommand{\hoch}[1]{$\, ^{#1}$}
\newcommand{\bea}{\begin{eqnarray}}
\newcommand{\eea}{\end{eqnarray}}
\newcommand{\ra}{\rightarrow}
\newcommand{\lra}{\longrightarrow}
\newcommand{\Lra}{\Leftrightarrow}
\newcommand{\ap}{\alpha^\prime}
\newcommand{\bp}{\tilde \beta^\prime}
\newcommand{\tr}{{\rm tr} }
\newcommand{\Tr}{{\rm Tr} }
\newcommand{\NP}{Nucl. Phys. }

\newcommand{\tamphys}{\it Center for Theoretical Physics\\
Texas A\&M University, College Station, Texas 77843}
\newcommand{\upenn}{\it Department of Physics and Astronomy\\
University of Pennsilvania, Philadelphia, PA 19104}
\newcommand{\auth}{M.J. Duff\hoch{\dagger 1}, P. Hoxha\hoch{\dagger}, H.
L\"u\hoch{\ddagger 2}, R.R. Martinez-Acosta\hoch{\dagger}
and C. N. Pope\hoch{\dagger 3}}

\begin{document}

\begin{flushright}
\hfill{CTP TAMU-43/97}\\
\hfill{UPR 0822-T}\\
\hfill{astro-ph/9712301}\\
\hfill{}
\end{flushright}

\vspace{20pt}

\begin{center}
{ \large {\bf A Lattice Universe from M-theory}}

\vspace{30pt}

\auth

\vspace{15pt}

\hoch{\dagger}{\tamphys}

\vspace{10pt}

\hoch{\ddagger}{\upenn}

\vspace{40pt}

\underline{ABSTRACT}
\end{center}

      A recent paper on the large-scale structure of the Universe
presented evidence for a rectangular three-dimensional lattice of
galaxy superclusters and voids, with lattice spacing $\sim 120~Mpc$,
and called for some ``hitherto unknown process'' to explain it. Here
we report that a rectangular three-dimensional lattice of intersecting
domain walls, with arbitrary spacing, emerges naturally as a classical
solution of M-theory.

{\vfill\leftline{}\vfill
\vskip  10pt
\footnoterule
{\footnotesize
        \hoch{1} Research supported in part by NSF Grant ????.
\vskip -12pt} \vskip   14pt
{\footnotesize
        \hoch{2} Research supported in part by DOE
Grant DE-FG02-95ER40893. \vskip       -12pt} \vskip 14pt
{\footnotesize
        \hoch{3} Research supported in part by DOE
Grant DE-FG03-95ER40917. \vskip       -12pt}}

\pagebreak
\setcounter{page}{1}

\section{Introduction}

      In a recent paper on the large-scale structure of the Universe
at the $100$-million parsec scale, Einasto {\it et al}
\cite{Einasto,Kirshner,Einasto1} report seeing hints of a network of
galaxy superclusters and voids that seems to form a three-dimensional
lattice with a spacing of about $120$ $h^{-1}$ $Mpc$ (where $h^{-1}$
is the Hubble constant in units of $100$ $km$ $s^{-1}$
$Mpc^{-1}$). These authors remark that ``If this reflects the
distribution of all matter (luminous and dark), then there must exist
some hitherto unknown process that produces regular structure on large
scales.''  In this paper we point out that a three-dimensional lattice
of orthogonally intersecting domain walls, with arbitrary lattice
spacing, naturally appears as a solution of the classical equations of
{\it $M$-theory}.

       Until recently, the best hope for an all-embracing theory that
would reconcile gravity and quantum mechanics was based on {\it
superstrings} \cite{Green}: one-dimensional objects whose vibrational
modes represent the elementary particles and which live in a universe
with ten spacetime dimensions, six of which are curled up to an
unobservably small size. Unfortunately, there seemed to be {\it five}
distinct mathematically consistent string theories and this was
clearly an embarrassment of riches if one is looking for a unique {\it
Theory of Everything}. In the last three years, however, it has become
clear that all five string theories may be subsumed by a deeper, more
profound, new theory called $M$-theory, which is reviewed in
Refs. \cite{Schwarz,Duff,Townsend,Banks}. $M$-theory is an {\it
eleven-dimensional} theory in which the extended objects are not
one-dimensional superstrings but rather two-dimensional objects called
{\it supermembranes} and five-dimensional objects called {\it
superfivebranes}. In the limit of low energies, $M$-theory is
approximated by eleven-dimensional supergravity \cite{Cremmer} which,
ironically enough, was the favourite candidate for superunification
before it was knocked off its pedestal by the 1984 superstring
revolution.

         Like string theory before it, $M$-theory relies crucially on
a {\it supersymmetry} which unifies bosons and fermions and indeed
eleven is the maximum spacetime dimension that supersymmetry
allows. The basic objects of $M$-theory are solutions of the
supergravity equations of motion which preserve the fraction $\nu=1/2$
of this supersymmetry.  In addition to the supermembrane and
superfivebrane there are also other solutions called {\it plane waves}
and {\it Kaluza-Klein monopoles} which also preserve $\nu=1/2$. Again,
seven of the eleven dimensions are assumed to be compactified to a
tiny size and, when wrapped around these extra dimensions, these
membranes, fivebranes, waves and monopoles appear as zero-dimensional
particles, one-dimensional strings or two-dimensional membranes when
viewed from the perspective of the physical four-dimensional
spacetime.  Of particular interest for the present paper will be the
membranes which act as domain walls separating one region of the
three-dimensional uncompactified space from another. Solutions
preserving a smaller fraction $\nu=2^{-n}$ of the supersymmetry may
then be obtained by permitting $n=2,3$ of the basic objects to
intersect orthogonally \cite{Pope}. In particular, a solution
describing $3$ domain walls intersecting orthogonally in three space
dimensions will, if it exists, preserve just $N=1$ of the maximum
$N=8$ supersymmetries allowed in four spacetime dimensions.

     Although astrophysicists and cosmologists have considered
topological defects \cite{Shellard} or {\it solitons} such as
monopoles, cosmic strings and domain walls as possible seeds for
galaxy formation, they have traditionally\footnote{See, however,
\cite{Elitzur}, where the periodic-like distribution of matter in
space \cite{Broadhurst} was noted in the context of stringy
topological defects.} eschewed the kind of solitons arising in
superstring theory \cite{Khuri}.  This was not without good
reason. They were interested in energy scales less than the order of
$10^{16}$ GeV typical of the energy at which the strengths of the
strong, weak and electromagnetic forces are deemed to converge in
Grand Unified Theories of the elementary particles. By contrast, the
solitons arising in string theory are gravitational in origin, having
as their typical energy scale the Planck energy of $10^{19}$ GeV which
corresponds to much too early an epoch in the history of the universe
to be relevant to galaxy formation. However, recent work by Witten
\cite{Witten} indicates that $M$-theory differs from traditional
superstring theory precisely in this respect: the phenomenologically
most favoured size of the eleventh compact dimension of $M$-theory is
such that all four forces converge at a common scale of $10^{16}$
GeV. Thus gravitational effects are much closer to home than
previously realised and topological defects in $M$-theory are likely
to be of greater cosmological significance than those of old-fashioned
string theory.

\section{The lattice universe}

       We now show that solutions describing three orthogonally
intersecting domain walls do indeed exist and that they may be
generalised to describe any number of walls so as to form a
three-dimensional lattice, with arbitrary lattice spacing.  (A review
of domain walls in $N=1$ $D=4$ supergravity can be found in
\cite{cvso}.)  We begin by considering a four-dimensional universe
with cartesian coordinates
$x^{\mu}=(x^{0},x^{1},x^{2},x^{3})=(t,x,y,z)$ whose gravitational
field is described by a metric tensor $g_{\mu\nu}(x)$ and which in
addition contains three massless scalar fields described by the
three-dimensional vector ${\vec \phi}(x)$ and a set of three $3$-form
potentials $C^{(\a)}_{\mu\nu\rho}, (\a=1,2,3)$. The Lagrangian is
given by the Einstein Lagrangian for gravity plus kinetic energy terms
for ${\vec \phi}(x)$ and $C^{(\a)}_{\mu\nu\rho}$,
\be
{\cal L}= \frac{1}{2\kappa^{2}}\sqrt{-g}\left[R
-\frac{1}{2}\partial^{\mu}{\vec \phi}\cdot\partial_{\mu}{\vec \phi}
-\frac{1}{48}\sum_{\a=1}^{3}e^{-{\vec c}_{\a}\cdot{\vec
\phi}}G^{(\a)}{}^{\mu\nu\rho\sigma}G^{(\a)}_{\mu\nu\rho\sigma}\right]
\la{Lagrangian}
\ee
where $\kappa^{2}=8\pi G$ and $G$ is Newton's constant and where
$G^{(\a)}{}_{\mu\nu\rho\sigma} \equiv
4\partial_{[\mu}C^{(\a)}{}_{\nu\rho\sigma]}$ are the $4$-form field
strengths associated with the $3$-form potentials.  In four spacetime
dimensions such $3$-forms do not correspond to propagating degrees of
freedom but may nevertheless give rise to non-trivial topological
effects which are akin to the presence of a cosmological constant
\cite{vann}.  The three vectors ${\vec c}_{\a}$ are constants
describing the interaction of the scalars with the field strengths and
satisfy the relations
\be
\vec c_\a \cdot \vec c_\b = 1 + 6\delta_{\a\b}\,
\la{c1}
\ee
A convenient choice is
\be
\vec c_{1}=(1,1 +{\scriptstyle{\sqrt{2}}},1-{\scriptstyle{\sqrt{2}}})
\ ,\quad
\vec c_{2}=(1-{\scriptstyle{\sqrt{2}}},1,1+{\scriptstyle{\sqrt{2}}})
\ ,\quad
\vec c_{3}=(1+{\scriptstyle{\sqrt{2}}},1-{\scriptstyle{\sqrt{2}}},1)\ .
\la{c2}
\ee
We shall shortly indicate the $M$-theoretic origin of the Lagrangian
(\ref{Lagrangian})
but first we provide the desired solution of the resulting field
equations:
\bea
\Box\vec\phi = -\ft1{48} \sum_\a \vec c_\a\, e^{-\vec
c_\a\cdot\vec\phi}\, (G^{(\a)})^2\ &,&\nn\\
\del_{\sst \mu}(e\, e^{-\vec c_\a\cdot\vec\phi}\, 
G^{(\a)\sst{\mu\nu\rho\sigma}})
= 0\ &,&\\
R_{\sst{\mu\nu}} = \ft12\del_{\sst \mu}\vec\phi\cdot \del_{\sst \nu}\vec\phi
 +\ft1{12} \sum_\a e^{-\vec c_\a\cdot\vec\phi}
\, ((G^{(\a)})^2_{\sst{\mu\nu}}
-\ft38 (G^{(\a)})^2 \, g_{\sst{\mu\nu}} )\ &.&\nn
\la{equations}
\eea
Let us make the following ansatz for the line element
\be
ds_4^{2}=g_{\mu\nu}dx^{\mu}dx^{\nu}=
(H_{1}H_{2}H_{3})^{-\frac{1}{2}}(-dt^{2}+
H_{1}dx^{2}+H_{2}dy^{2}+H_{3}dz^{2}),
\la{metric}
\ee
for the scalars
\be
{\vec \phi}=-\frac{1}{2}\sum_{\a=1}^3\vec c_\a\, \log~H_{\a},
\la{scalars}
\ee
and for the $4$-form field strengths
\be
G^{(\a)}=H_{\b}H_{\gamma}\partial_{\a}H_{\a}{}^{-1}
dt \wdg dx \wdg dy \wdg dz\ ,
\la{fourforms}
\ee
where $\a$ $\b$ and $\gamma$ are all different, and take their values
from the set $1,2$ and $3$. Here the functions have the coordinate
dependences $H_{1}(x),H_{2}(y),H_{3}(z)$.  Substituting into the field
equations resulting from the Lagrangian above, we find that all are
satisfied provided the functions $H_{\a}$ are harmonic:
\be
\frac{\partial^{2}H_{1}}{\partial x^{2}}=
\frac{\partial^{2}H_{2}}{\partial y^{2}}=
\frac{\partial^{2}H_{3}}{\partial z^{2}}=0\ .
\la{H}
\ee
These harmonic conditions can be satisfied by taking $H_{\a}$ to
have the form
\be
H_{1}(x)= 1+\sum_{a=1}^{a={\cal N}}M_{a}|x-x_{a}|
\la{walls}
\ee
and similarly for $H_{2}(y)$ and $H_{3}(z)$. From the form of the
metric, we see that if $H_{2}$ and $H_{3}$ are temporarily taken to be
unity, then the function $H_{1}$ describes a stack of ${\cal N}$
parallel domain walls lying in the $(y,z)$ plane at the locations
$x=x_{a}$ whose mass per unit area $M_{a}$ is also equal to the
$3$-form charge.  Similarly, the functions $H_{2}$ and $H_{3}$ by
themselves can describe stacks of domain walls in the $(x,z)$ and
$(x,y)$ planes respectively.  When all three functions are of the
general form given above, the solution describes the triple
intersection of domain walls lying in the three planes orthogonal to
$x$, $y$ and $z$ axes, respectively. Actually, the expressions for the
$H_{\a}$ are not harmonic everywhere, since they have delta-function
singularities at the locations of the domain walls:
\be
\frac{\partial^{2}H_{1}}{\partial x^{2}}=2\sum_{a=1}^{a={\cal
N}} M_{a} \delta(x-x_{a})\ ,\qquad \hbox{etc.}
\la{sources}
\ee
This may be remedied by including in the field equations a source
term\footnote{In eleven-dimensions, duality relates the elementary
singular ``electric'' membrane, which requires a source term in the
$D=11$ supergravity field equations, to the solitonic non-singular
``magnetic'' fivebrane, which solves the source-free equations. It
thus appears that the process of dualisation can eliminate the need
for a source! This confusion is presumably a consequence of the
inadequacy of $D=11$ supergravity to capture the full essence of
M-theory.}  for each membrane in a way that preserves the
supersymmetry \cite{Khuri}.

\section{Higher dimensional origins}

The Lagrangian (\ref{Lagrangian}) we have been considering is obtained
from the bosonic sector of eleven-dimensional supergravity
\cite{Cremmer} which contains a metric $g_{MN}$ and a $3$-from
potential $A_{MNP}$ where $M=0,1,\ldots ,10$.  For simplicity we
assume that the seven compactified dimensions have the topology of a
seven-dimensional torus. The three scalars ${\vec \phi}$ are a subset
of the $7$ scalars coming from the diagonal components of the internal
metric, while the three $4$-form field strengths
$G^{(\a)}{}_{\mu\nu\rho\sigma}$ appear after dualising a subset of the
$0$-form field strengths which arise via a generalised Scherk-Schwarz
type of dimensional reduction in which tensor potentials coming both
from the eleven-dimensional metric $g_{MN}$ and $3$-form potential
$A_{MNP}$, are allowed to depend linearly on the compactification
coordinates.  By retracing these steps it is possible to re-express
our four-dimensional solution as an eleven-dimensional solution.

To see the origin of dilaton vectors satisfying the relations
(\ref{c1}), we note that the vectors $\vec c_\a$ are just the
negatives of the dilaton vectors for the $0$-form field strengths that
we are dualising in order to obtain the $4$-forms $G^{(\a)}$.  The
$0$-form field strengths arise from the Scherk-Schwarz reduction of
$1$-form fields; in other words, from reductions where the axionic
scalar potentials for the $1$-form field strengths are allowed linear
dependences on the compactification coordinates.  Details of such
Scherk-Schwarz reductions and subsequent dualisations may be found in
Ref. \cite{brgpt,clpst,Pope}. There are two sources of $0$-form field
strengths, namely those coming from the Scherk-Schwarz reduction of
the antisymmetric tensor in $D=11$, and those coming from the
Scherk-Schwarz reduction of the Kaluza-Klein vectors coming from the
metric in $D=11$.  These $0$-forms, denoted by $F_0^{(ijk\ell)}$ and
${\cal F}_0^{(ijk)}$ respectively in \cite{Pope}, have associated
dilaton vectors $\vec a_{ijk\ell}$ and $\vec b_{ijk}$, given by
\be
\vec a_{ijk\ell} =\vec f_i +\vec f_j +\vec f_k +\vec f_\ell -\vec g\ ,
\qquad \vec b_{ijk} =-\vec f_i + \vec f_j +\vec f_k\ ,
\ee
where
\be
\vec g\cdot \vec g = 7\ ,\qquad \vec g\cdot\vec f_i = 3\ ,\qquad
\vec f_i\cdot\vec f_j =1+2\delta_{ij}
\ee
in $D=4$.  Note that the indices $i,j,\ldots$ range over the $7$
compactification coordinates $z_i$ here, starting with $i=1$ for the
reduction step from $D=11$ to $D=10$.  It is now easy to see that one
can indeed find three dilaton vectors that satisfy the conditions
(\ref{c1}).  There are many different combinations that will work, but
they are all equivalent, up to index relabellings, to the two choices
\be
\vec c_1 = \vec a_{1234}\ ,\qquad \vec c_2 =\vec a_{1567}\ ,
\qquad \vec c_3 = \vec b_{125}\ .\label{3vecs}
\la{choice1}
\ee
or
\be
\vec c_1 = \vec a_{1234}\ ,\qquad \vec c_2 = \vec b_{235}\ ,\qquad \vec
c_3 = \vec b_{345}\ .
\la{choice2}
\ee
Note that for the first choice (\ref{choice1}) we must necessarily
choose two $0$-forms that originate from $F_4=dA_{3}$ in $D=11$, and
one coming from the metric.  In particular we must necessarily use
terms which originate, in ten-dimensional type IIA language, from both
the NS-NS and R-R sectors of the theory.  This means that our
intersecting membranes are solutions of the type IIA theory or
M-theory, but not of the heterotic or type I strings. However, it is
also possible, as in choice (\ref{choice2}) to choose the dilaton
vectors $\vec c_{\alpha}$ such that the $D=4$ solution originates only
from $D=10$ NS-NS sector. As such it could equally well be regarded as
a solution of the $D=10$ heterotic or type I string.

    The reason why we chose to work with the dual formulation with
$4$-form field strengths, rather than the original $0$-form field
strengths coming from the Scherk-Schwarz reduction, is that the former
description is necessary, in the framework of a four-dimensional
theory itself, if one wants to have the possibility of multiple
membranes in a stack along each coordinate axis.  The reason for this
is that a $0$-form field strength term of the form $-\ft12 M^2\,
e^{\vec c\cdot \vec \phi}$ in the four-dimensional Lagrangian would
lead to an equation of motion that required the associated harmonic
function $H$ to have slope $\pm M$, and hence we could only have a
solution of the form $H=1+ M\, |x-x_0|$. In order to have
straight-line segments of different slopes, as is required in
(\ref{walls}) in order to describe multiple membranes, it is necessary
that the slope can take different values in different regions, and so
it must arise as an arbitrary constant of integration (related to the
the electric charge of $F_4$) rather than as a fixed, given parameter
(i.e $M$) in the Lagrangian.

    From the higher-dimensional point of view, the restriction of
needing to work in the dualised formulation with $4$-form field
strengths in $D=4$ is actually an artificial one.  The reason for this
is that the $0$-form field strengths in the four-dimensional theory
themselves arise from the generalised Scherk-Schwarz reductions of
standard degree $1$ or higher field strengths in higher dimensions.
In these reductions, unlike in standard Kaluza-Klein reductions, the
associated potentials are allowed linear dependences on certain
compactifying coordinates.  In our example above, for instance, the
$0$-form field strength $F_0^{(1234)}$ arises from a generalised
Scherk-Schwarz reduction step from $D=8$ to $D=7$, where the axion
$A_0^{(123)}$ is reduced according to $A_0^{(123)}(x,z_4) \rightarrow
A_0^{(123)}(x) + M\, z_4$. This gives the 0-form field strength term
$-\ft12 M^2\, e^{\vec a_{1234}\cdot\vec\phi}$ in $D=7$, and lower,
dimensions.  However the constant $M$ is itself an integration
constant from the viewpoint of the theory in $D=8$, and so provided we
trace our four-dimensional solution back to $D=8$ or a higher
dimension, the parameter $M$ of the four-dimensional Lagrangian in the
0-form formulation can be understood as nothing but a free integration
constant.  Thus again, it can take different values in different
regions, and so the possibility of having multiply-charged solutions
is regained.

     In view of the above, it is therefore useful to re-interpret our
intersecting membrane solutions back in higher dimensions.  This is
easily done by retracing the steps of the Kaluza-Klein and
Scherk-Schwarz reductions.  Taking our first choice of field strengths
corresponding to (\ref{choice1}), we find that in $D=10$ the solution
becomes
\bea
ds_{10}^2 &=& (H_1\, H_2)^{-1/4}\, H_3^{-1/8}
\Big(-dt^2 + H_1\, (dx^2+dz_3^2
+dz_4^2) + H_2\, (dy^2 + dz_6^2 + dz_7^2) \nn\\
&&\qquad\qquad\qquad\qquad\quad + H_3\, dz^2 + H_1\, H_3
dz_2^2 + H_2\, H_3\, dz_5^2\Big )\ ,\nn\\
e^{\phi_1} &=& (H_1\, H_2)^{-1/2}\, H_3^{3/4}\ ,\label{d10sol}\\
F_3^{(1)}&=& -\del_x H_1\,  dz_2\wedge dz_3 \wedge dz_4
-\del_y H_2\, dz_5\wedge dz_6 \wedge dz_7
\ ,\qquad {\cal F}_2^{(1)} =-\del_z H_3\, dz_2\wedge dz_5\ ,\nn
\eea
where $\phi_1$, the first component of the fields $\vec \phi$ in
$D=4$, is the dilaton of the type IIA theory.\\
The supersymmetry transformation parameter of this solution satisfies the
following relations
\bea
( 1 - \Gamma_{023\hat5\hat6\hat7} )\, \epsilon &=& 0 \ , \nn\\
( 1 - \Gamma_{013\hat2\hat3\hat4} )\, \epsilon &=& 0 \ , \label{susy1}\\
( 1 - \Gamma_{012\hat3\hat4\hat6\hat7} )\, \epsilon &=& 0 \ , \nn
\eea
where $\epsilon$ is given in terms of a constant spinor $\epsilon_0$:
\be
\epsilon = (H_1 H_2)^{-1/16}\, H_3^{-1/32}\, \epsilon_0\ , \label{eps1}
\ee
where $\Gamma_{\sst M}$ are the $D=10$ Dirac matrices, and
$\Gamma_{\sst{M_1\cdots M_n}}=
\Gamma_{[{\sst M_1}}\cdots \Gamma_{{\sst M_n}]}$.  (Note that the
explicit numerical indices 0, 1, 2 and 3 refer to the four-dimensional
spacetime, while $\hat1$, $\hat2$, {\it etc}. refer to the reduction
steps $i=1,2$ {\it etc}. from $D=11$ to $D=10,9$ {\it etc}.)  
Therefore, the above
solution preserves the fraction $\nu=1/8$ of the supersymmetry.  The
solution (\ref{d10sol}) can be viewed as a non-standard intersection of
two NS-NS 5-branes and one D6-brane.  The harmonic functions here depend
on the relative transverse coordinates, which differ from the situation in
standard intersections, where the harmonic functions depend only on the
overall transverse space.  The two 5-brane charges are both carried by
the NS-NS 3-form field strength $F_3^{(1)}$, and the D6-brane charge
is carried by the R-R 2-form field strength ${\cal F}_2^{(1)}$.  The
solution with vanishing D6-brane charge was also obtained in
\cite{khuri,bbj,gkt}.

     Going back one step further, to $D=11$, the solution becomes
\bea
ds_{11}^2 &=& (H_1 \, H_2)^{-1/3}\, \Big(-dt^2 + H_1\, (dx^2+dz_3^2+dz_4^2)
+H_2\, (dy^2+dz_6^2 +dz_7^2) \nn\\
&&+ H_3\, dz^2
+ H_1\, H_3\, dz_2^2 + H_2\,H_3\, dz_5^2 + H_1\, H_2\, H_3^{-1}\,
(dz_1+ \del_z\, H_3\, z_5\, dz_2)^2 \Big)\ ,\nn\\
F_4 &=& -\del_x H_1\, dz_1\wedge dz_2\wedge dz_3 \wedge dz_4
-\del_y H_2\, dz_1\wedge dz_5\wedge dz_6 \wedge dz_7 \ .\label{d11sol}
\eea
This solution can be viewed as a non-standard intersection of two
M5-branes and one NUT (again with the harmonic functions depending on
the relative transverse coordinates rather than those of the overall
transverse space).

    It is also possible to choose the dilaton vectors $\vec
c_{\alpha}$ such that the $D=4$ solution originates from $D=10$ NS-NS
sector as with (\ref{choice2}).  By retracing the steps of the
Kaluza-Klein and Scherk-Schwarz reductions we find that in $D=8$ the
solution becomes
\bea
ds_{8}^2 &=& H_3^{-1/6} \Big(-dt^2 + H_1\, dx^2 + H_2\, dy^2 + H_3\,
dz^2 + H_1\,H_3\, dz_4^2 \nn\\
&&\qquad\quad + H_2\,H_3\, dz_5^2 + dz_6^2 + dz_7^2\Big) \ , \nn\\
\vec \phi &=& - \ft12 \vec a_{123}\, \log\, H_1 - \ft12 \vec b_{23}\, \log\,
H_2 - \ft12 \vec b_3\, \log\, H_3 \ , \label{d8sol}\\
F_1^{(123)} &=& \del_x H_1\,dz_4\ ,\qquad {\cal F}_1^{(23)} = \del_y H_2\,
dz_5 \ , \qquad {\cal F}_2^{(3)} = \del_z H_3\, dz_4 \wedge dz_5\ , \nn
\eea
where
\bea
\vec a_{123} &=& \Big(\, 1 ,\quad \ft3{\sqrt 7} ,\, 2\sqrt{\ft37}\, \Big)
\ , \nn\\
\vec b_{23} &=& \Big(\, 0 , -\ft4{\sqrt 7} ,\, 2\sqrt{\ft37}\, \Big) \ ,
\label{vecs8}\\
\vec b_3 &=& \Big(\, 0 ,\quad 0\, ,\, -\sqrt{\ft73}\, \Big) \ . \nn
\eea
$D=8$ is the highest dimension where the metric remains diagonal in the
oxidation process.\\
The solution in $D=10$ takes the form
\bea
ds_{10}^2 &=& H_1^{-1/4} \Big(-dt^2 + H_1\, dx^2 + H_2\, dy^2 + H_3\,
dz^2 + H_1\,H_3\, dz_4^2 \nn\\
&&\qquad\quad + H_2\,H_3 \, dz_5^2  + H_1\,H_2\,H_3^{-1} \,(dz_3 + z_4
\, \del_z H_3\, dz_5)^2 + dz_6^2+ dz_7^2 \nn\\
&&\qquad\quad + H_1\, H_2^{-1} (dz_2 - z_3\, \del_y H_2 \, dz_5)^2 \Big)
\ , \nn\\ e^{\phi_1} &=& H_1^{-1/2}\ ,\label{d10sol2}\\
F_3^{(1)} &=& \del_x H_1\, dz_4\wedge (dz_2 - z_3\, \del_y H_2 \, dz_5)
\wedge (dz_3 + z_4\, \del_z H_3\, dz_5)\ . \nn
\eea
The $D=10$ supersymmetry transformation parameter satisfies the conditions
\bea
( 1 + \Gamma_{1\hat2\hat3\hat4} )\, \epsilon &=& 0 \ , \nn\\
( 1 + \Gamma_{3\hat3\hat4\hat5} )\, \epsilon &=& 0 \ , \label{susy2}\\
( 1 - \Gamma_{2\hat2\hat3\hat5} )\, \epsilon &=& 0 \ , \nn
\eea
where
\be
\epsilon = H_1^{-1/16}\, \epsilon_0\ . \label{eps2}
\ee
Thus, the fraction of the supersymmetry preserved by this solution is
also $\nu=1/8$.  This can regarded as a solution of the type 1, heterotic
or type IIA string. In the latter case, the solution can be further
oxidised to $D=11$, where it becomes becomes
\bea
ds_{11}^2 &=& H_1^{-1/3} \Big(-dt^2 + H_1\, dx^2 + H_2\, dy^2 + H_3\,
dz^2 + H_1\,H_3\, dz_4^2 \nn\\
&&\qquad\quad + H_2\,H_3\,  dz_5^2+ +dz_6^2 + dz_7^2 +
H_1\,H_2^{-1}\, (dz_2 - z_3\,  \del_y H_2\, dz_5)^2 \nn \\
&&\qquad\quad + H_1\,H_2\,H_3^{-1}\, (dz_3+ z_4\,  \del_z H_3 \, dz_5)^2 +
H_1\, dz_1^2 \Big) \ , \nn\\
F_4 &=&  \del_x H_1\, dz_4\wedge (dz_2 - z_3\, \del_y H_2 \, dz_5)
\wedge (dz_3 + z_4\, \del_z H_3 \, dz_5)\wedge dz_1
\ . \label{d11sol2}
\eea
In both $D=10$ and $D=11$ dimensions, the solutions can be viewed as
non-standard intersections of a 5-brane with two NUTs.  In $D=10$, the
5-brane carries the NS-NS charge.

          In this section, we have considered two examples of
non-standard intersections in $D=10$ string theory or $D=11$ M-theory
that can give rise to a four-dimensional lattice universe.  The first
example is intrinsic to M-theory, while the second example can
alternatively be embedded in the heterotic string.  More solutions can
be obtained by invoking the T-duality of the type IIA and type IIB
theories.  We shall not enumerate such examples here.

\section{Conclusions}

   The worldwide web solution we have presented is admittedly an
idealisation, and by itself would not yet satisfy hard-nosed
astrophysicists. First, it represents a static universe with an
unbroken supersymmetry, whereas in the real world the universe is
expanding and supersymmetry (if it exists at all) is a broken
symmetry. In fact these two features are intimately related. The
reason we were able to find a stable static three-dimensional lattice,
rather than one which is collapsing under its own gravity or one whose
tendency to collapse is overwhelmed by expansion, is precisely because
of the famous ``no-static-force'' phenomenon \cite{Khuri} of
supersymmetric vacua which saturate a Bogomol'nyi-Prasad-Sommerfield
bound between the mass and the charge.  The mutual gravitational
attraction due to gravity $g_{\mu\nu}$ and the massless scalar fields
${\vec \phi}$ is exactly cancelled by a repulsion due the the
$3$-forms $C^{(\a)}{}_{\mu\nu\rho}$, which act in many ways like a
cosmological constant. (In writing this, of course, we are acutely
aware that the introduction of a cosmological constant in order to
obtain a static rather than expanding universe was, on his own
admission, Einstein's ``greatest blunder''.)  A more realistic
description must therefore await a satisfactory explanation of how
M-theory breaks supersymmetry and, unfortunately, this remains
M-theory's biggest unsolved problem. Of course, our solution is but
one of many solutions of M-theory. We make no apology for this. The
lattice structure is no more a {\it prediction} of M-theory than the
Friedman-Robertson-Walker cosmology is a prediction of General
Relativity.

     The intersecting domain wall configuration is one where the
regions of high density are concentrated on the faces of the lattice
cubes. One might ask whether there are also solutions where the
regions of high density are concentrated on the edges of the cube
(intersecting strings associated with $3$-form field strengths) or on
the vertices of the cube (point singularities associated with $2$-form
field strengths). The experimental data do not sharply distinguish
between these possibilities, since they depend on the overdensity
$\delta\rho/\rho$ chosen in making the statistical analysis.  We have
found no such intersecting string solutions and consider their
existence unlikely. There are solutions describing any number of
isolated points which may, in particular, be chosen to lie on a cubic
lattice. These are $M$-theoretic generalisations of the well-known
Papapetrou-Majumber solutions of general relativity. Once again the
mutual attraction due to gravity and scalars is exactly cancelled by a
repulsion due to the $1$-form potentials.

        Another reason why astrophysicists might object is that domain
walls whose mass per unit area is too great are ruled out
experimentally.  In the supersymmetric idealisation presented here,
the mass per unit area is also a free parameter depending on the
vacuum expectation values of the scalar fields which, for simplicity
of presentation, we have arbitrarily set equal to zero. Once again,
the actual value of these scalar expectation values must await a
resolution of the supersymmetry-breaking problem.

      Finally, we are aware that with the success of inflationary
models of the universe, topological defects have fallen out of favour
as the mechanism for galaxy formation \cite{Turok}, although the issue 
remains controversial \cite{Jesus}. In any event, it is
difficult to see how inflation {\it alone} could account for the
three-dimensional cubic lattice reported by Einasto et
al. \cite{Einasto}. To the best of our knowledge, all attempts to fit
this lattice structure data with data on the cosmic microwave
background have been based on entirely ad hoc assumptions on the
initial spectrum of density perturbations. See, for example,
\cite{Einasto2}. Moreover, the phenomenology of the kinds of defect
appearing in M-theory has yet to be scrutinized. Consequently, in
spite of the idealised nature of our solution we hope to have shown
that M-theory is indeed a rich source of possible explanations for
``hitherto unexplained phenomena'' and in particular allows for
three-dimensional lattice cosmologies.

\section{Acknowledgements}

We are grateful to John Barrow and Alexei Starobinsky for useful
correspondence.


\begin{thebibliography}{99}

\bibitem{Einasto}
J. Einasto, M. Einasto, S. Gottlober, V. Muller, V. Saar, A.A.
Starobinsky, E. Tago, D. Tucker, H. Andernach and P. Frisch,
\newblock {\sl A $120~Mpc$ periodicity in the three-dimensional
distribution of galaxy superclusters},
\newblock  Nature, {\bf 385} (1997) 139.

\bibitem{Kirshner}
R. Kirshner,
\newblock {\sl The Universe as a lattice},
\newblock  Nature, {\bf 385} (1997).

\bibitem{Einasto1}
J. Einasto,
\newblock{\sl Has the universe a honeycomb structure?},
\newblock astro-ph/9711320

\bibitem{Green}
M.B. Green, J.H. Schwarz and E. Witten,
{\sl Superstring theory},
Cambridge University Press (1987).

\bibitem{Schwarz}
J.H. Schwarz,
\newblock {\sl The power of M-theory},
\newblock Phys. Lett. {\bf B367} (1996) 97, hep-th/9510086.

\bibitem{Duff}
M.J. Duff,
\newblock {\sl  $M$-theory (the theory formerly known as strings)},
\newblock  I. J. M. P {\bf A11} (1996) 5623, hep-th/9608117.

\bibitem{Townsend}
P.K. Townsend,
\newblock {\sl Four lectures on M-theory},
\newblock hep-th/9612121.

\bibitem{Banks}
T. Banks,
\newblock {\sl Matrix Theory},
\newblock hep-th/9710231.

\bibitem{Cremmer}
E. Cremmer, B. Julia and J. Scherk,
\newblock {\sl Supergravity theory in $11$ dimensions},
\newblock Phys. Lett. {\bf B76} (1978) 409.

\bibitem{Pope}
H. L\"u, C.N. Pope, T.A. Tran and K.W. Xu,
\newblock {\sl Classification of $p$-branes, NUTs, waves and intersections},
\newblock to appear in Nucl. Phys. {\bf B}, hep-th/9708055.

\bibitem{Elitzur}
S. Elitzur, A. Forge and E. Rabinovici,
\newblock{\sl Some global aspects of string compactifications},
Nucl. Phys. {\bf B359} (1991) 581.

\bibitem{Khuri}
M. J. Duff, R.R. Khuri and J.X. Lu,
{\sl String solitons},
Phys. Rep. {\bf 259} (1995) 213, hep-th/9412184

\bibitem{Shellard}
A. Vilenkin and E.P.S. Shellard,
\newblock{\sl Cosmic strings and other topological defects},
\newblock Cambridge University Press 1994.

\bibitem{Witten}
E. Witten,
\newblock {\sl Strong coupling expansion of Calabi-Yau compactification},
\newblock Nucl. Phys. {\bf B471} (1996) 135, hep-th/9602070.

\bibitem{Broadhurst}
T. J. Broadhurst, R. S. Ellis, D. C. Koo and A. S. Szalay,
\newblock {\sl Large scale distribution of galaxies at the galactic
poles},
\newblock Nature {\bf 343} (1990) 726.

\bm{cvso} M. Cvetic and H.H Soleng, {\sl Supergravity domain walls},
Phys. Rep. {\bf 282} (1997) 159, hep-th/9604090.

\bibitem{vann}
M.J. Duff and P. van Nieuwenhuizen,
\newblock {\sl Quantum inequivalence of different field representations},
\newblock Phys. Lett. {\bf B94} (1980) 179.

\bibitem{brgpt} E. Bergshoeff, M. de Roll, M.B. Green, G. Papadopoulos
and P.K. Townsend, {\it Duality of type II 7 branes and 8 branes},
Nucl. Phys. {\bf B470} (1996) 113, hep-th/9601150.

\bibitem{clpst} P.M. Cowdall, H. Lu, C.N. Pope, K.S. Stelle and
P.K. Townsend, {\sl Domain walls in massive supergravities},
Nucl. Phys. {\bf B486} (1997) 49, hep-th/9608173.

\bm{khuri} R.R. Khuri, {\sl A comment on string solitons}, Phys. Rev.
{\bf D48} (1993) 2947, hep-th/9305143.

\bm{bbj} K. Behrndt, E. Bergshoeff and B. Janssen, {\sl Intersecting
D-branes in ten and six dimensions}, Phys. Rev. {\bf D55} (1997) 3785,
hep-th/9604168.

\bm{gkt} J.P. Gauntlett, D.A. Kastor and J. Traschen, {\sl Overlapping
branes in M-theory}, Nucl. Phys. {\bf B478} (1996) 544, hep-th/9604179.

\bibitem{Turok}
U. Seljak, U-L. Pen and N. Turok,
\newblock{\sl Power spectra in global defect theories of cosmic
structure formation},
Phys. Rev. Lett. {\bf 79} (1997) 1615.

\bibitem{Jesus}
J. Pando, D. Valls-Gabaud and Li-Zhi-Fang,
\newblock{\sl Evidence for scale-scale correlations in the cosmic 
microwave background radiation},
astro-ph/9810165.

\bibitem{Einasto2}
F. Atrio-Barandela, J. Einasto, S. Gottlober, V. Muller, A.A.
Starobinsky,
\newblock {\sl A built-in scale in the initial spectrum of density
perturbations: evidence from cluster and CMB data},
\newblock  JETP Lett. {\bf 66} (1997) 397, astro-ph/9708128.

\end{thebibliography}
\end{document}